\documentstyle[epsfig]{mn}
\newif\ifAMStwofonts
                                                                                                                                                     
\ifoldfss
  \ifCUPmtlplainloaded \else
    \NewTextAlphabet{textbfit} {cmbxti10} {}
    \NewTextAlphabet{textbfss} {cmssbx10} {}
    \NewMathAlphabet{mathbfit} {cmbxti10} {} 
    \NewMathAlphabet{mathbfss} {cmssbx10} {} 
  \fi
  \ifAMStwofonts
    \ifCUPmtlplainloaded \else
      \NewSymbolFont{upmath} {eurm10}
      \NewSymbolFont{AMSa} {msam10}
      \NewMathSymbol{\upi}     {0}{upmath}{19}
      \NewMathSymbol{\umu}     {0}{upmath}{16}
      \NewMathSymbol{\upartial}{0}{upmath}{40}
      \NewMathSymbol{\leqslant}{3}{AMSa}{36}
      \NewMathSymbol{\geqslant}{3}{AMSa}{3E}

      \let\leq=\leqslant 
       
    \fi
  \fi
\fi 
\ifnfssone
  \newmathalphabet{\mathit}
  \addtoversion{normal}{\mathit}{cmr}{m}{it}
  \addtoversion{bold}{\mathit}{cmr}{bx}{it}
  \newmathalphabet{\mathbfit} 
  \addtoversion{normal}{\mathbfit}{cmr}{bx}{it}
  \addtoversion{bold}{\mathbfit}{cmr}{bx}{it}
  \newmathalphabet{\mathbfss} 
  \addtoversion{normal}{\mathbfss}{cmss}{bx}{n}
  \addtoversion{bold}{\mathbfss}{cmss}{bx}{n}
  \ifAMStwofonts
    \ifCUPmtlplainloaded \else
      \UseAMStwoboldmath
      \makeatletter
      \new@mathgroup\upmath@group
      \define@mathgroup\mv@normal\upmath@group{eur}{m}{n}
      \define@mathgroup\mv@bold\upmath@group{eur}{b}{n}
      \edef\UPM{\hexnumber\upmath@group}
      \new@mathgroup\amsa@group
      \define@mathgroup\mv@normal\amsa@group{msa}{m}{n}
      \define@mathgroup\mv@bold\amsa@group{msa}{m}{n}
      \edef\AMSa{\hexnumber\amsa@group}
      \makeatother
      \mathchardef\upi=''0\UPM19
      \mathchardef\umu=''0\UPM16
      \mathchardef\upartial=''0\UPM40
      \mathchardef\leqslant=''3\AMSa36
      \mathchardef\geqslant=''3\AMSa3E

      \let\leq=\leqslant 

    \fi
  \fi
\fi 
                                                                                                                                                     
\ifnfsstwo
  \DeclareMathAlphabet{\mathbfit}{OT1}{cmr}{bx}{it}
  \SetMathAlphabet\mathbfit{bold}{OT1}{cmr}{bx}{it}
  \DeclareMathAlphabet{\mathbfss}{OT1}{cmss}{bx}{n}
  \SetMathAlphabet\mathbfss{bold}{OT1}{cmss}{bx}{n}
  \ifAMStwofonts
    \ifCUPmtlplainloaded \else
      \DeclareSymbolFont{UPM}{U}{eur}{m}{n}
      \SetSymbolFont{UPM}{bold}{U}{eur}{b}{n}
      \DeclareSymbolFont{AMSa}{U}{msa}{m}{n}
      \DeclareMathSymbol{\upi}{0}{UPM}{``19}
      \DeclareMathSymbol{\umu}{0}{UPM}{``16}
      \DeclareMathSymbol{\upartial}{0}{UPM}{``40}
      \DeclareMathSymbol{\leqslant}{3}{AMSa}{``36}
      \DeclareMathSymbol{\geqslant}{3}{AMSa}{``3E}

      \let\leq=\leqslant 

    \fi
  \fi
\fi 
                                                                                                                                                     
\ifCUPmtlplainloaded \else
  \ifAMStwofonts \else 
    \def\upi{\pi}
    \def\umu{\mu}
    \def\upartial{\partial}
  \fi
\fi

\title{The non-Gaussian Cold Spot in WMAP: significance, morphology and foreground contribution}

\author[M. Cruz et al.]{M. Cruz,$^{1,2}$\thanks{e-mail:
cruz@ifca.unican.es}
M. Tucci,$^3$
E. Mart\'{\i}nez-Gonz\'alez,$^1$
and P. Vielva$^1$ \\
$^1$IFCA, CSIC-Univ. de Cantabria, Avda. los Castros, s/n, 39005-Santander,
Spain\\
$^2$Dpto. de F\'{\i}sica Moderna, Univ. de Cantabria, Avda. los Castros, 
s/n, 39005-Santander, Spain\\
$^3$Astrophysics Group, The Blackett Laboratory, Imperial College, London SW7 2AZ UK}

\date{Accepted  Received ; in original form }

\begin{document}


\maketitle
                                                                                                                                                    

\begin{abstract}

The non--Gaussian cold spot in the 1--year WMAP data, described in Vielva et al. and Cruz et al., is analysed in detail in the 
present paper. First of all, we perform a more rigorous calculation of the significance of the non--zero kurtosis detected in WMAP maps by Vielva 
et al. in wavelet space, mainly generated by \emph{the Spot}. We confirm the robustness of that detection, since the probability of obtaining 
this deviation by chance is 0.69\%.
Afterwards, the morphology of \emph{the Spot} is studied by applying Spherical Mexican Hat Wavelets with different ellipticities. The shape of 
\emph{the Spot} is found to be almost circular.
Finally, we discuss if the observed non--Gaussianity in wavelet space
can arise from bad subtracted foreground residues in the WMAP maps. 
We show that the flat frequency dependence of \emph{the Spot} cannot be explained 
by a thermal Sunyaev--Zeldovich effect.
Based on our present knowledge of Galactic foreground emissions, we conclude that
the significance of our detection is not affected by Galactic
residues in the region of \emph{the Spot}. Considering different Galactic foreground estimates, 
the probability of finding such a big cold spot in Gaussian simulations is always below 1\%.

\end{abstract}

\begin{keywords}
methods: data analysis - cosmic microwave background
\end{keywords}

\section{Introduction}

The Cosmic Microwave Background (CMB) is currently the most valuable tool to study the origin of the universe. This radiation
was emitted $\approx300000$ years after the Big--Bang and is the most ancient electromagnetic radiation that we can observe. 
It offers us information about the conditions in which it was emitted, soon after the recombination of the atoms that made the universe 
transparent to photons. 

The most accepted theory of the early universe is the standard inflationary scenario. The CMB presents small temperature anisotropies.
According to the standard model, these anisotropies represent a homogeneous and isotropic Gaussian random field on the celestial sphere. 
However, non-standard inflation (Bartolo et al. 2004) and topological defect 
models (Turok \& Spergel 1990 and Durrer 1999) predict different non-Gaussian features in the CMB sky.

The most precise all--sky measurement of the CMB is up to date the Wilkinson Microwave Anisotropy Probe (WMAP, Bennett et al. 2003a). 
The analysis of the WMAP team did not find any evidence of deviations from Gaussianity of the cosmological signal 
in the 1--year data (Komatsu et al. 2003).
Nevertheless in the last years many scientists studied the Gaussianity of these data and found non--Gaussian
signatures or asymmetries using different methods: namely low multipole alignment statistics (de Oliveira--Costa et al. 2004a, Copi et al. 2004, 2005, 
Schwarz et al. 2004, Land \& Magueijo 2005a,b,c, Bielewicz et al. 2005, Slosar \& Seljak 2004); phase correlations (Chiang et al. 2003, 
Coles et al. 2004); hot and cold spot analysis (Larson \& Wandelt 2004, 2005); local curvature methods (Hansen et al. 2004, Cabella et al. 2005); 
correlation functions (Eriksen et al. 2004a, 2005, Tojeiro et al. 2005); structure alignment statistics (Wiaux et al. 2006);
multivariate analysis (Dineen \& Coles 2005); Minkowski functionals (Park 2004, Eriksen et al. 2004b); gradient and dispersion analyses 
(Chyzy et al. 2005) or wavelets (Vielva et al. 2004, Mukherjee \& Wang 2004, Cruz et al. 2005, McEwen et al. 2005).
Also upcoming analyses with new tools, such as scalar statistics on the sphere, (Monteser\'{\i}n et al. 2005), or steerable filters,
(Wiaux et al. 2005) may find some more non--Gaussian features. 

\begin{figure}
  \begin{center}
    \includegraphics[width=5cm,angle=-90]{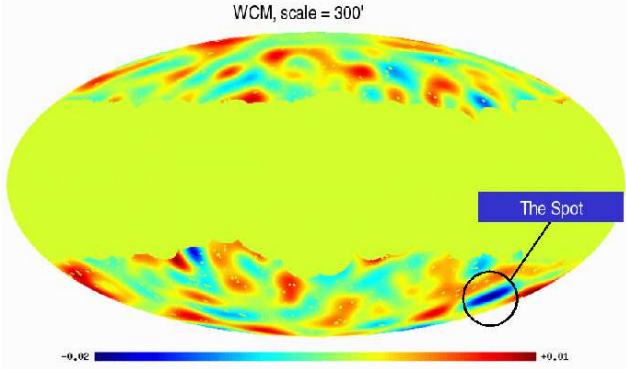}
  \end{center}
  \caption{The combined and foreground cleaned Q--V--W WMAP map after
    convolution with the SMHW at scale $R_{9}$. The position of \emph{the
    Spot} is marked.}
  \label{fig:SPOT}
\end{figure}

In this paper we focus on the Gaussianity studies provided by Vielva et al. (2004) and Cruz et al. (2005), hereafter V04 and C05. 
These are based on the Spherical Mexican Hat Wavelet (SMHW) technique which allows us to perform a multiscale analysis and to localise the non--Gaussian 
signatures. 
In V04 an excess of kurtosis was detected convolving the data with the SMHW at scales between $3\degr$ and $5\degr$. The deviation from 
Gaussianity presented an upper tail probability of 0.38\% considering all the sky and 0.11\% in the southern hemisphere, analysing both Galactic hemispheres
independently. 
Performing an analysis of the spots, C05 found that this deviation 
was due exclusively to a cold spot at galactic coordinates ($b = -57^\circ, l = 209^\circ$), called \emph{the Spot} (see Fig.~\ref{fig:SPOT}).
Assuming the Gaussian hypothesis, the upper tail probability of having such a spot at a wavelet scale of $5\degr$ is 0.18\%. 
These detections were confirmed by other papers, also based on wavelets. Mukherjee \& Wang (2004) and McEwen et al. (2005) 
recalculated the kurtosis analysis of V04 obtaining similar results. 
Cay\'on, Jin \& Treaster (2005) applied higher criticism statistics to maps in wavelet space. They detect a deviation from Gaussianity, 
finding that some pixels of \emph{the Spot} are responsible for it.

One of the aims of this paper is to re--analyse carefully the robustness of V04 results. In this paper 15 different wavelet scales and
two estimators, kurtosis and skewness were considered. 
Could it be that looking at that many estimators, the deviation from Gaussianity just happened by chance?

Anyway, the origin of non-Gaussian signatures is in general still not clear and controversial.
Many scientists just believe that their findings are due to a deficient subtraction of the Galactic foreground emissions, 
(e.g. Chiang \& Naselsky 2004 and Tojeiro et al. 2005).
Some others (e.g. Eriksen et al. 2005) argue that foregrounds are unlikely to explain their results.

Considering the non--Gaussianity in wavelet space and related to the
very big and cold spot, several possibilities were discussed in V04 and C05, 
namely systematics, Galactic foregrounds, thermal Sunyaev--Zeldovich effect (Sunyaev \& Zeldovich 1970, hereafter SZ effect), topological defects 
or gravitational effects (Rees \& Sciama 1968, Mart{\'\i}nez--Gonz{\'a}lez et al. 1990a, b).

A better knowledge of \emph{the Spot}'s shape can help us to find out its origin.
For example, 
\emph{the Spot} could be explained by considering topological defect models, as suggested in C05. In this context, if \emph{the Spot} presents
circular symmetry, it could have been generated by a texture. 
Finding out the shape of \emph{the Spot}, can also inform us about a hypothetical gravitational potential that could be generating the 
non-Gaussian emission. Tomita (2005) suggests a local second--order gravitational effect as a possible origin of \emph{the Spot}. 
These possibilities will be discussed in future papers.

The easiest  explanation assuming the widely accepted Gaussian hypothesis would be a bad foreground subtraction of the data. Hence it should be
the first one to be tested.
V04 and C05 performed several tests to discard the foregrounds as a cause of the detection. The most important test is the
frequency dependence of \emph{the Spot}, since the Galactic foregrounds have a strong frequency dependence. The amplitude and the number of pixels 
of \emph{the Spot} have been shown to have a constant frequency dependence, as expected from a cosmological origin.
Nevertheless, some recent works seem to contradict these conclusions.
Recently, Liu \& Zang (2005) analysed the non--Gaussianity results obtained with spherical wavelets, concluding that the most possible source of 
the non--Gaussian departures, are residual foregrounds. These conclusions do not agree with the results presented in V04 and C05. 
Also Coles (2005) suggest that Galactic foregrounds could be behind our and other non-Gaussian features.

Other different explanations are suggested for the large scale asymmetries and the non-Gaussian behaviour found in wavelet space.
Jaffe et al. (2005a) assume an anisotropic cosmological model, namely the Bianchi type VII$_h$. The best fit Bianchi model introduces 
an additional CMB component, which presents a swirl pattern. Looking at this pattern, a hot and a cold spot can be seen close to the center of the swirl. 
The cold spot of the best fit model matches \emph{the Spot}. Subtracting the swirl pattern, the significance of the kurtosis deviation and other 
large scale anomalies are largely reduced. Nevertheless a later work, Jaffe et al. (2005b), seems to indicate the incompatibility of the WMAP data
with extended Bianchi models including a dark energy term.
Freeman et al. (2005) points out a possible incorrect dipole correction as a plausible explanation for some of the observed asymmetries. 
This hypothesis is under examination but since \emph{the Spot} has a much lower scale than the dipole, we do not expect to be able to explain it in this way.

Considering the relevance of the issue and the present controversial
debate, the main part of this work is dedicated to investigate the origin of non--Gaussianity found in wavelet space and, in
particular a possible foreground contribution to \emph{the Spot}. Different techniques of
foreground subtraction are taken into account.

Summarising, first we will describe shortly the data and simulations in section $\S$2; the study of the robustness of the non--Gaussian
detection in wavelet space will be discussed in section $\S$3; the
morphology of \emph{the Spot} in section $\S$4; the foreground tests
are presented in $\S$5, and our conclusions in section $\S$6.

\section{Data and Simulations}

We will use the 1-year WMAP data, available in the Legacy Archive
for Microwave Background Data Analysis (LAMBDA) web page\footnote{http://lambda.gsfc.nasa.gov/product/map}.
The WMAP data are given at five frequency-bands, namely K--band (22.8 GHz, one  receiver), 
Ka--band (33.0 GHz, one  receiver), Q--band (40.7 GHz, two  receivers), V--band (60.8 GHz, two  receivers) and W--band (93.5 GHz, four  receivers). 
Also the foreground cleaned maps for the Q, V, and W channels are available. 
The main analysis in V04 and C05 was performed using the combined, foreground cleaned Q--V--W map (hereafter WCM; see Bennett et al. 2003a).
The use of this map is recommended by the WMAP-team. It is a noise weighted linear combination of the receivers at frequencies where CMB is the 
dominant signal. 

However some other foreground cleaned maps have been suggested by (Bennett et al. 2003b), Tegmark et al. (2003), Eriksen et al. (2004c) or 
Patanchon et al. (2005). In section $\S$5 we will describe in more detail some of the different available foreground cleaned maps.

In order to test the Gaussianity of the data, we use the 10000 Gaussian simulations described in V04 and C05.

The Hierarchical, Equal Area and iso-Latitude Pixelization (HEALPix, G\'{o}rski et al. 2005) with resolution parameter $N_{side}=256$, is used in all maps.
In order to mask the contaminated Galactic region, we apply the kp0 mask (Bennett et al. 2003b) plus some extended masks for the convolution 
with wavelets (see V04).

\section{Robustness of the non--Gaussian detection in wavelet space}

It is not trivial at all to determine how significant the non--Gaussian detections are. Almost all analyses define \emph{the Spot} as a highly significant
deviation from Gaussianity, but the given significance differs. Our aim in this section is to find the probability of obtaining our observation by chance, 
assuming that the Gaussian hypothesis is true. This probability is called $p$--value and will give us the robustness of our findings. The $p$--value should 
not be confused with the upper tail probability, which is the cumulated probability above the observed value.
Before entering in the discussion, let us summarise the obtained results in previous works.

In the first work, V04 compared the skewness and the kurtosis of the data with 10000 Gaussian simulations, at 15 wavelet scales, between 13.7 and 1050 arcmin.

The deviation of the kurtosis with respect to the Gaussian hypothesis was found to have an upper tail probability of $0.38\%$ at wavelet scale $R_9 = 5\degr$.
Considering the Galactic hemispheres separately, the lowest upper tail probability appears in the southern one and is $0.11\%$ at scale $R_7 = 3.33\degr$, whereas
the northern hemisphere is compatible with Gaussianity.

Afterwards C05 performed a spot analysis using the same scales and simulations as in V04. The considered estimators were: number of spots,
number of maxima and minima, and number of pixels above or below a given threshold (hot or cold area). At the same scales where the kurtosis excess
was found in V04 and at high thresholds, the cold area shows up very low upper tail probabilities and the number of cold spots was just one.
The lowest probability, $0.18\%$, was obtained at scale $R_8 = 4.17\degr$.  In this way the cold spot located at ($b = -57^\circ, l = 209^\circ$) was discovered to be 
the one responsible for the non-­Gaussian detection.

C05 compared the area of \emph{the Spot} to the biggest spot of each simulation, but only for scales $R_8$ and $R_9$. At these scales all the calculated 
probabilities were lower than $0.65\%$ (see Table 2 of C05), being the lowest one $0.18\%$ at scale $R_9$.

All these upper tail probabilities have been calculated counting how many simulations have higher or equal values of our estimator than the data
at \emph{one particular scale}.
Since several scales and estimators were considered, these upper tail probabilities do not tell us the $p$--value, i.e. probability of obtaining our 
observation by chance assuming that the Gaussian hypothesis is true.
We have to bear in mind that considering a large enough number of estimators and scales, 
the probability of finding non--Gaussian features in any Gaussian simulation would increase significantly.

We should therefore review our calculations. 
As for the area of \emph{the Spot} only two scales were considered, one could think that this would be the best estimator to calculate the $p$--value.
In all other cases we considered 15 scales and several estimators
However in the analyses after the detection of the non-zero kurtosis, the choice of the scales was conditioned by this first finding. C05 considered
only two scales for the area of \emph{the Spot} because there the kurtosis deviation was the largest.
Hence we have to test the robustness with the first detection, the deviation of the kurtosis.
This deviation occurs mainly at 3 scales considering all the sky, and at 4 scales considering only the southern hemisphere.

McEwen et al. (2005) calculated the $p$--value for the all--sky case, using a very conservative approach.
They search through 1000 Gaussian simulations to determine the number of maps that have an equivalent or greater deviation than the maximum deviation 
found in the data. The obtained $p$--value was $4.97\%$.

This approach does not take into account that the deviation from Gaussianity occurs at several consecutive scales.
As already mentioned, in the all--sky case scales $R_7$, $R_8$ and $R_9$ present a significant deviation from Gaussianity. 
The highest upper tail probability of these three scales is $0.67\%$.

In order to know how likely it is to find such a detection by chance, we have to answer the following question: How many of the 10000 Gaussian 
simulations, show up a higher or equal deviation in any three consecutive scales, and in any of the two estimators, kurtosis and skewness?
This number will give us the $p$--value.
In other words, we have to find how many simulations show skewness or kurtosis values with upper or lower tail probabilities lower 
than $0.67\%$ in any three consecutive scales. The $p$--value we obtain is much lower than the previous one and turns out to be $1.91\%$.

Anyway the most significant upper tail probability of the kurtosis was not obtained in the all--sky case.
Considering the northern and southern hemispheres separately the deviation is highly significant at 4 scales, namely $R_6 = 2.5\degr$, $R_7$, $R_8$ and 
$R_9$. The highest upper tail probability of these four scales is $0.55\%$. 
Now we search through the Gaussian simulations to find how many simulations show skewness or kurtosis values with upper or lower tail probabilities 
lower than $0.55\%$ in any four consecutive scales and in any of the two hemispheres. The $p$--value is now $0.69\%$ proving
that the kurtosis deviation from Gaussianity is robust.

Dividing the sky into two hemispheres the significance is higher, since the deviation in the kurtosis is due to a localised spot. Anyway we
cannot continue dividing the sky in an unlimited number of regions, as the sample variance increases. In our case we are still far away from this
limit because each hemisphere contains a large enough number of pixels. 

However we should remind that calculating the exact significance of any non-Gaussian analysis is not easy. 
In our present approach the calculation is more rigorous
but with an a-posteriori interpretation, i.e. given the fact that the data deviate from Gaussianity at three consecutive scales, we calculate the probability
of obtaining a similar deviation in the simulations. The most important conclusion is that \emph{the Spot} remains statistically robust, independently of the 
chosen significance test.

\section{Morphpology of the Spot}
\begin{table*}
   \begin{center}
         \caption{Minimum temperature ($\mu$K) of \emph{the Spot} after convolution of the WCM with the ESMHW at scale $R_{8}=4.17\degr$.} 	 
         \begin{tabular}{|c|c|c|c|c|c|c|c|c|c|}
	  orientation  & $\rho$ = & 0.125 & 0.250 & 0.375 & 0.500 & 0.625 & 0.750 & 0.875 & 1.000 \\
	 \hline
	   1 && -8.14 & -9.77 & -11.04 & -12.89 & -14.55 & -15.62 & -16.09 & -16.06 \\
	   2 && -5.49 & -8.95 & -11.07 & -12.62 & -13.90 & -14.93 & -15.65 & -16.06 \\
	   3 && -3.60 & -6.50 &  -8.38 & -10.59 & -12.56 & -14.19 & -15.37 & -16.06 \\
	   4 && -2.37 & -4.42 &  -7.28 & -10.44 & -12.91 & -14.61 & -15.62 & -16.06 \\
	   5 && -5.67 & -8.45 & -11.18 & -13.38 & -15.00 & -15.86 & -16.16 & -16.06 \\
	   6 && -3.35 & -5.82 &  -9.46 & -12.63 & -14.83 & -16.04 & -16.33 & -16.06 \\
	 \end{tabular}
	 \label{table:minESMHW}
   \end{center}
\end{table*}
As our detection was performed using the SMHW, which has axial symmetry, \emph{the Spot} is expected to be almost circularly symmetric
when observed in wavelet space. By convolving the map 
with the SMHW, we are in fact amplifying all those underlying signals whose shape is similar to that of the SMHW, whereas other signals are lowered.
In particular the zero level is removed with the SMHW since it is a compensated wavelet.

However the shape of \emph{the Spot} could be not completely symmetric. In order to characterise \emph{the Spot}, we should construct an
Elliptical Mexican Hat Wavelet on the Sphere (ESMHW) to find out if there is any preferred direction which amplifies even more \emph{the Spot}.
Information about the shape of \emph{the Spot} would help us to determine its possible origin.
The ESMHW should have the same properties as the SMHW but with an elliptic section instead of a circular one. 

The Elliptical Mexican Hat Wavelet on the plane, is a generalisation of the symmetric 2D Mexican Hat Wavelet and therefore is proportional to the
Laplacian of an \emph{elliptical} Gaussian function,

\begin{eqnarray}
\label{eq:2DEMHW}
\Psi(x_{1},x_{2},a,b) & \propto & \Delta \exp{\left[-\left(\frac{x_{1}^2}{2a^2}+\frac{x_{2}^2}{2b^2}\right)\right]}
\\
           \nonumber  & \propto & \left[{a}^2 + {b}^2 - \left(\frac{{b}^2}{{a}^2}{x_{1}}^2 + \frac{{a}^2}{{b}^2}{x_{2}}^2\right)\right] \\
                      & \nonumber & \nonumber \times \exp{\left[-\left(\frac{x_{1}^2}{2a^2}+\frac{x_{2}^2}{2b^2}\right)\right]}
\end{eqnarray}
with two different scales $a$ and $b$, $0 < b \leq a < \infty$.
In order to compare with the SMHW, we define the scale $R$, and the axial ratio $\rho$, as follows
\begin{eqnarray}
\label{eq:R, x}
   R & \equiv & \sqrt{ab}   \\
   \rho & \equiv & \frac {b}{a} , (0 \leq \rho \leq 1). 
\end{eqnarray}

To obtain the Ellpitical Mexican Hat on the Sphere, we perform the stereographic projection (Mart\'{\i}nez-Gonz\'alez et al. 2002) 
defined by $(\mathbf{x}) \mapsto (\theta,\phi)$

\begin{eqnarray}
\label{eq:projection}
x_{1} & = & 2\tan \frac{\theta}{2} \cos \phi
\\
x_{2} & = & 2\tan \frac{\theta}{2} \sin \phi,
\end{eqnarray}
where ($\theta,\phi$) are the polar coordinates on the sphere.
The distance on the tangent plane is given by $y$ that is related to the polar 
angle ($\theta$) through:

\begin{equation}
\label{eq:y}
      y\equiv \sqrt{x_{1}^2+x_{2}^2} = 2\tan \frac{\theta}{2}.
\end{equation}
The Jacobian of the transformation is 
\begin{equation}
\label{eq:Jacobian}
J  =  \cos^{-4} \frac{\theta}{2} = {\left[1+{\left(\frac{y}{2}\right)}^2\right]}^2.
\end{equation}
Hence we obtain
\begin{equation}
\label{eq:ESMHW}
\Psi_S(\theta,\phi,a,b)  =  \frac{16J}{\sqrt{2\pi N}} \left[{a}^2+{b}^2-{k}^2{y}^2\right]\exp{\left(-{r}^2{y}^2/2\right)},
\end{equation}
where $k$, $r$ are defined as follows
\begin{eqnarray}
\label{eq:kr}
k & = & \rho \sqrt{ 1-\left(1-{\rho}^4\right) \sin^2 \phi }
\\
r & = & b^{-1} \sqrt{ 1-\left(1-{\rho}^2\right) \cos^2 \phi},
\end{eqnarray}
and $N$ is the normalisation constant which has been chosen such that $\int d\theta d\phi \sin \theta \Psi_S^2(\theta,\phi,a,b) = 1$.
%
%
%

Then we have convolved the ESMHW defined in Eq.~(\ref{eq:ESMHW}) with the combined and foreground cleaned map (WCM).
We have chosen eight axial ratios $\rho$ for scale $R_{8}=4.17\degr$, and six equally spaced orientations of the
ESMHW centered on \emph{the Spot}. In Table~\ref{table:minESMHW} we report the minimum temperature of \emph{the Spot} after the
convolution with the different ESMHW.

The value of the minimum temperature depends on how much the underlying signal is amplified. If the spatial distribution of the signal matches the
shape of the considered wavelet, the amplification is higher.
From  Table~\ref{table:minESMHW} we see that the minimum temperatures for low axial ratios are much lower than the others, for all orientations.
The minimum temperature of \emph{the Spot} is reached at axial ratios 1 or 0.875, being the differences between the six orientations small.
These results indicate that \emph{the Spot} is essentially circular. Similar conclusions can be
obtained for scales $R_{7}$ and $R_{9}$.

\section{Foreground contribution to the Non--Gaussian Spot}

Our aim in this section is to study possible foreground contributions to the non--Gaussian cold spot. 
In particular, we will focus on SZ and Galactic foregrounds. Because of their different
spectral behaviour respect to CMB, we expect to obtain relevant
information from a frequency analysis of \emph{the Spot}. 

It is important to remind that, since the convolution with a wavelet is a linear
operation, the frequency dependence of a component with no spatial variation in its spectrum, is the
same in wavelet space as in real space. Considering that we will perform our analysis in a a small patch of the sky, 
we assume that there are no significant variations in the spectral index even for synchrotron emission.

\subsection{The Sunyaev--Zeldovich effect}

The SZ effect causes a decrement in the temperature, for frequencies lower than 217 GHz. This effect occurs when
CMB photons cross hot electron gas inside a galaxy cluster, suffering inverse Compton scattering, where the electrons transfer energy to the
CMB photons. Hence several low frequency photons are promoted to higher frequencies. For the WMAP frequencies, the SZ effect has a negative
contribution and has to be considered as a possible source for a very cold spot as the one we are studying.

We already know that the huge size and very low temperature of \emph{the Spot} are very difficult to explain with the SZ effect (C05). In 
addition there are no observed clusters in this region according to the Abell, Corwin \& Olowin (1989) catalogue. 
However it is worth checking if the frequency dependence of the SZ
effect is compatible with the data, since an unobserved or dark cluster could be present in the mentioned region of the sky.

\begin{figure}
  \begin{center}
    \includegraphics[width=84mm]{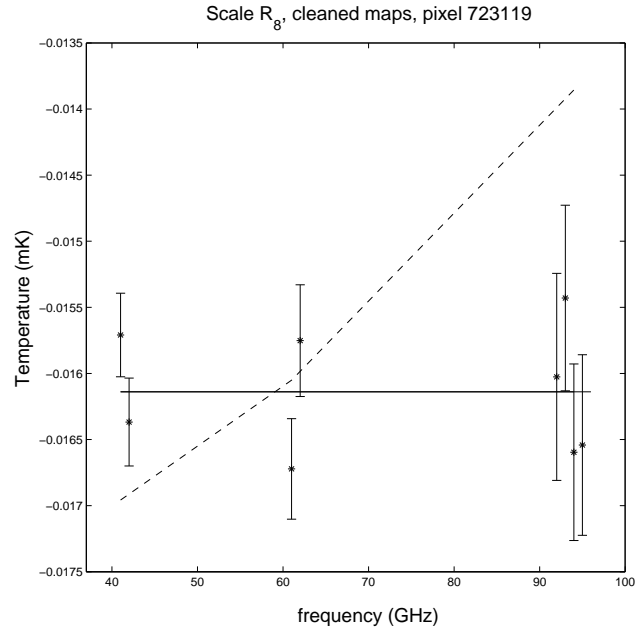}
  \end{center}
  \caption{The temperature at the center of \emph{the Spot} for channels Q, V and W at
    Scale $R_{8}$. CMB (solid line) and SZ (dashed line) are fitted to the data.
    The data at the same frequency have been slightly offset in abscissa for readability.}
  \label{fig:Q_V_W_R8}
\end{figure}

In Fig.~\ref{fig:Q_V_W_R8} we plot the temperature at the center of \emph{the Spot} ($b = -57^\circ, l = 209^\circ$)
at wavelet scale $R_{8}$.
Since the convolution with wavelets involves many pixels in this
region, only one pixel is representative for the entire spot. In the
plot we consider the values corresponding to the eight foreground cleaned, individual assembly data:
two for the Q and V bands and four for
the W--band. We estimate the
error bars for these temperatures performing 1000 noise simulations for
each of the eight receivers, and convolving them with the SMHW at scale
$R_{8}$. The standard deviation of the 1000 wavelet coefficients at
the chosen pixel plus the corresponding calibration error is the
error bar for each receiver.

Fig.~\ref{fig:Q_V_W_R8} does not show any evidence for a frequency
dependence of the temperature (the same is observed with the area
of \emph{the Spot}). In fact, assuming no Galactic contamination in \emph{the
Spot}, we find that the CMB fit of the data, i.e.
$\Delta T_{CMB} = -16.09 \pm 0.16 \mu$K, has a reduced
chi--square of 1.00. In the figure, we also represent the frequency
spectrum for thermal SZ effect, that is in thermodynamic temperature
\begin{equation}
\label{eq:freqDependence}
\Delta T_{SZ}(\nu) = \bigg[\frac{h \nu}{k_B}\coth
\left(\frac{h\nu}{2k_BT_0}\right)-4T_0\bigg]y_{c},
\end{equation}
where $T_0$ is the CMB temperature and $y_{c}$ the Compton parameter.
It can be clearly seen that
the SZ alone does not fit the data (the reduced chi--square is
9.12). If we perform a CMB+SZ fit, the Compton parameter and hence the SZ
effect are compatible with zero. 
We can conclude that these results strongly discard the SZ hypothesis, whereas
the data fit very well to a constant CMB value.

\subsection{Relevance of Galactic foregrounds in the region of the Spot}
\label{s1}

In this section we discuss in detail the contribution of Galactic
foregrounds in the region of \emph{the Spot}, both in real and in 
wavelet space. 
Some authors have suggested that residues of Galactic foregrounds are supposed to be the most likely source for
the non--Gaussian behaviour observed in CMB maps obtained
from the WMAP data (Chiang \& Naselsky 2004;
Liu \& Zhang 2005; Tojeiro et al. 2005). This is because their intensity
distribution in the sky is strongly non--Gaussian, and there
are still big uncertainties on the Galactic emission at microwave
wavelengths. Hence, our aim is to check if this hypothesis is reliable.
We focus only on \emph{the Spot} region because, as C05 pointed out, deviations from
Gaussianity detected by all--sky statistics in wavelet space (like
kurtosis, number and area of cold spots) are mainly produced by this
region. Excluding \emph{the Spot} area from those analyses makes
the WMAP data fully consistent with Gaussianity. The individual
peculiarity of \emph{the Spot} is stressed in C05 for its dimension in the
sky and for its very ``cold'' temperature.

To this purpose, we use templates of the different Galactic components
in \emph{the Spot} region, extrapolated to the WMAP frequencies. For the {\bf
free--free} emission we consider the H$\alpha$ map of Finkbeiner (2003),
with magnitudes corrected by the Galactic reddening E(B--V) map
provided by Schlegel, Finkbeiner \& Davis (1998). The data are converted from Rayleigh to
antenna temperature using Eq.\,11 of Dickinson, Davies \& Davis (2003), assuming an
electron temperature of 7000K. As {\bf synchrotron} template we use
the Rhodes/HartRAO 2326--MHz radio survey (Jonas et al. 1998). At this
frequency the contribution of free--free is expected to be low but not
negligible and it is subtracted using the H$\alpha$ map appropriately
scaled. Then, the Rhodes survey is extrapolated to higher frequencies by a
power law with spectral index $-3$. This value is in agreement with
the average synchrotron spectral index observed in WMAP maps
(Bennett et al. 2003b, Bernardi et al. 2004). It is also the average spectral
index found between 0.408 and 2.326\,GHz in \emph{the Spot} region. Respect
to the Haslam map, used by WMAP team, the Rhodes survey has the advantage of
a higher resolution and frequency, providing a more reliable template
for synchrotron at cosmological frequencies. Finally we consider the
full--sky map of {\bf thermal dust} at $100\mu$K generated by
Schlegel, Finkbeiner \& Davis (1998), extrapolated to microwaves by the best two--component
dust model (model 8) found by Finkbeiner et al. (1999), hereafter referred as FDS model.

In the last years some works seem to support the existence of spinning
dust emission at frequencies around 10--20 GHz
(de Oliveira-Costa et al. 2004b, Watson et al. 2005). However this
emission could provide a significant contribution to the WMAP data only for the K and Ka bands, whereas it quickly
decreases at higher frequencies, becoming negligible for the V and W bands (Draine \& Lazarian 1999). 
Therefore, we neglect this emission because, even if a contribution of spinning dust was present in the Spot region 
at K and Ka bands, it would not affect the following analysis.

At the moment, the non--Gaussianity tests have been performed on the
combined and foreground cleaned Q--V--W map (WCM). In this map the Galactic signal is removed by fitting
simultaneously a set of external foreground templates to the
residual Q--V--W maps (see Bennett et al 2003b). However, in the literature,
other techniques have been employed to subtract Galactic foregrounds
from WMAP maps and to produce clean CMB maps. Therefore, using different and
independent maps can give us a further test to verify that the
non--Gaussian deviations observed by V04 and C05 are not dependent
on the foreground--subtraction technique. For example, Bennett et al (2003b)
produced a clean map by a weighted internal linear combination (ILC)
of the five frequency WMAP maps.
In this way, the resulting CMB map does not rely on external foreground templates, but assumes that there are no spatial variations
in the frequency spectrum of each component and introduces complex noise properties.
Following the same idea,
Tegmark et al. (2003) provided a CMB map by an internal linear combination with
weights depending on frequency, latitude and angular scales. They
found a good agreement with the ILC map at large angular scales and no
excess of power at $\ell<100$ due to foregrounds. However also the TCM map presents complex noise properties (Eriksen et al 2004c).
Tegmark et al. (2003)
derived also a Wiener filtered map. This map was built for visualization purposes, as a best guess of how CMB looks like, and is not an unbiased map. 

Finally, Bennett et al (2003b) provide a description of how to build MEM
foreground estimates, even if they suggest not to apply the MEM
solutions directly to CMB analyses because of their complicated noise
and signal correlations. 
We obtain a MEM cleaned map, adding the eight not cleaned individual assembly maps, for channels Q, V and W, and subtracting 
the MEM foreground estimates for all these channels.

The WCM and ILC maps are not completely independent, as the latter has been
used in the foreground subtraction of the WCM map. 
Therefore, we shall focus on WCM and TCM maps, i.e. the two independent and most
reliable CMB maps (leaving the ILC, Wiener and MEM solutions only as
complementary test in wavelet space, where the signal to noise ratio is higher).

\subsubsection{The Spot in real space}
\label{ss1}

\begin{figure*}
\begin{center}
\includegraphics[width=14cm]{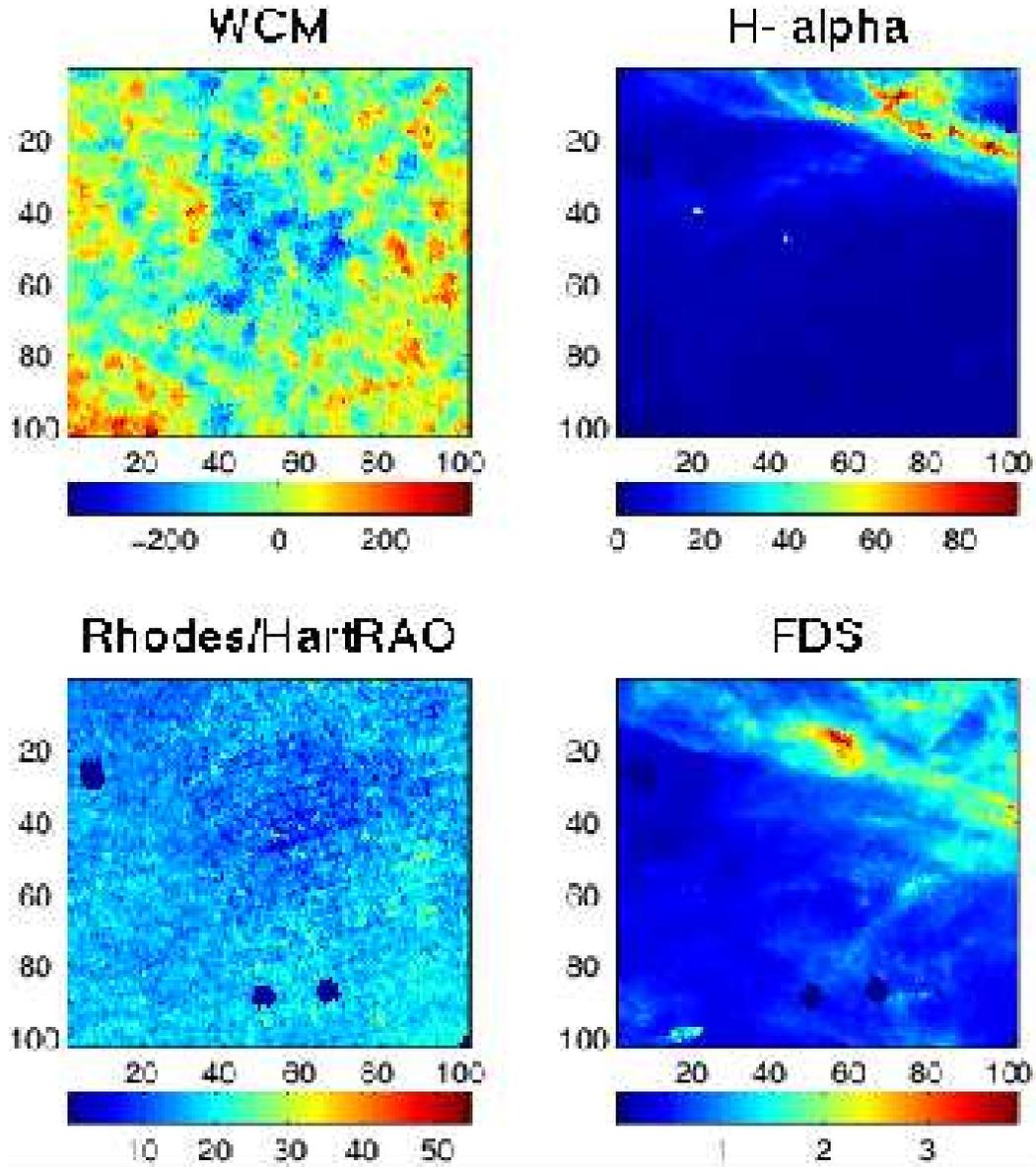}
\end{center}
\caption{Image showing an azimuthal projection of a $22\degr \times 22\degr$ patch from HEALPix maps with resolution nside = 256, 
in the region of \emph{the Spot}. 
From top to bottom and left to right, we have: the WCM map where \emph{the Spot} can be seen; the H-alpha map; the Rhodes/HartRAO 2326--MHz radio survey;
and the FDS dust template. The three foreground templates are scaled to channel Q, where the foreground contribution is higher than in the V and W bands.
The maps are in $\mu$K units and the labels on the axes are in pixels of size $13.3 \times 13.3$ arcmin and the y--axis is oriented in the Galactic 
North-South direction.}
\label{fig:SpotImage}
\end{figure*}
\begin{figure*}
\includegraphics[width=84mm, height=84mm]{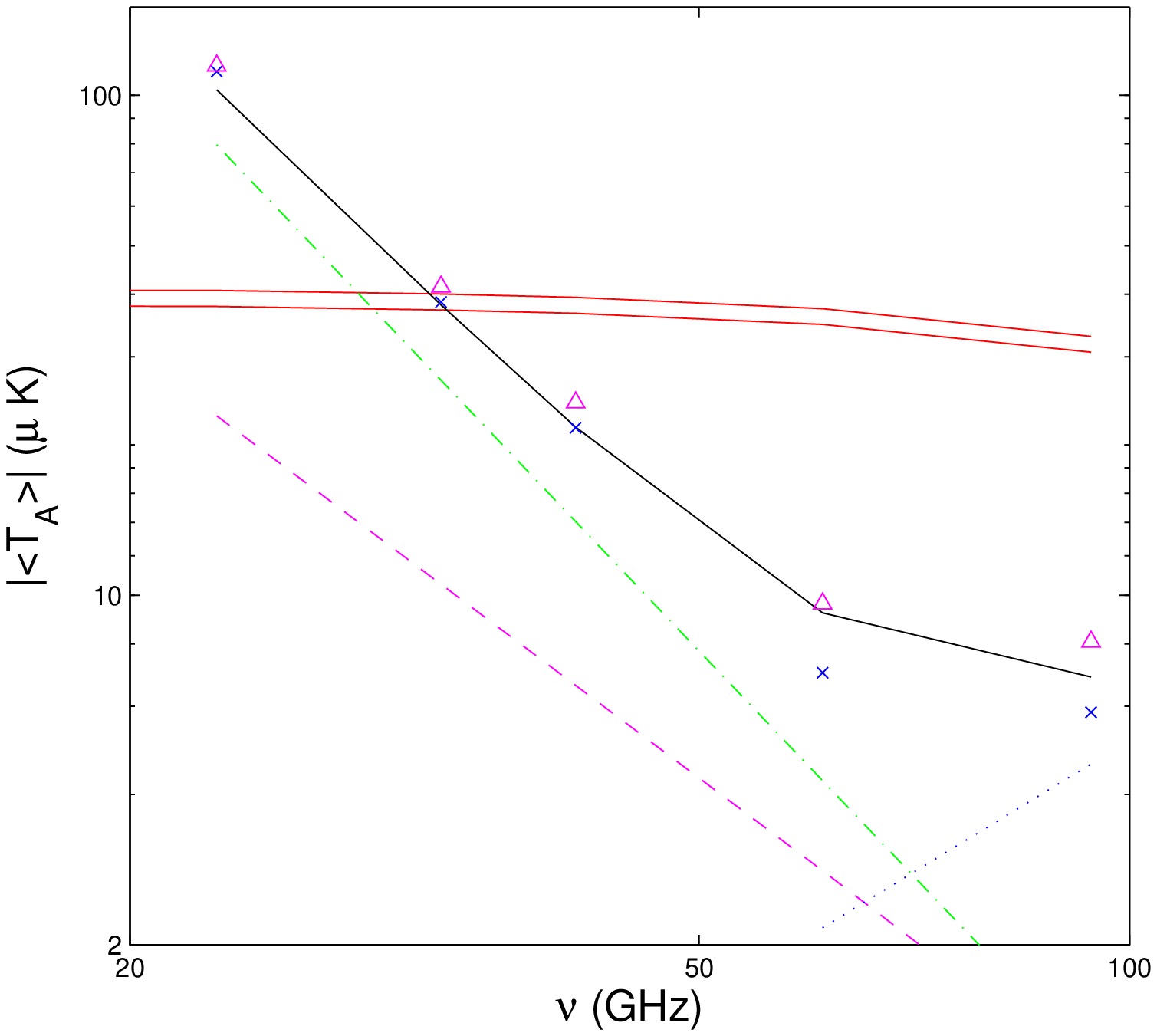}
\includegraphics[width=84mm, height=84mm]{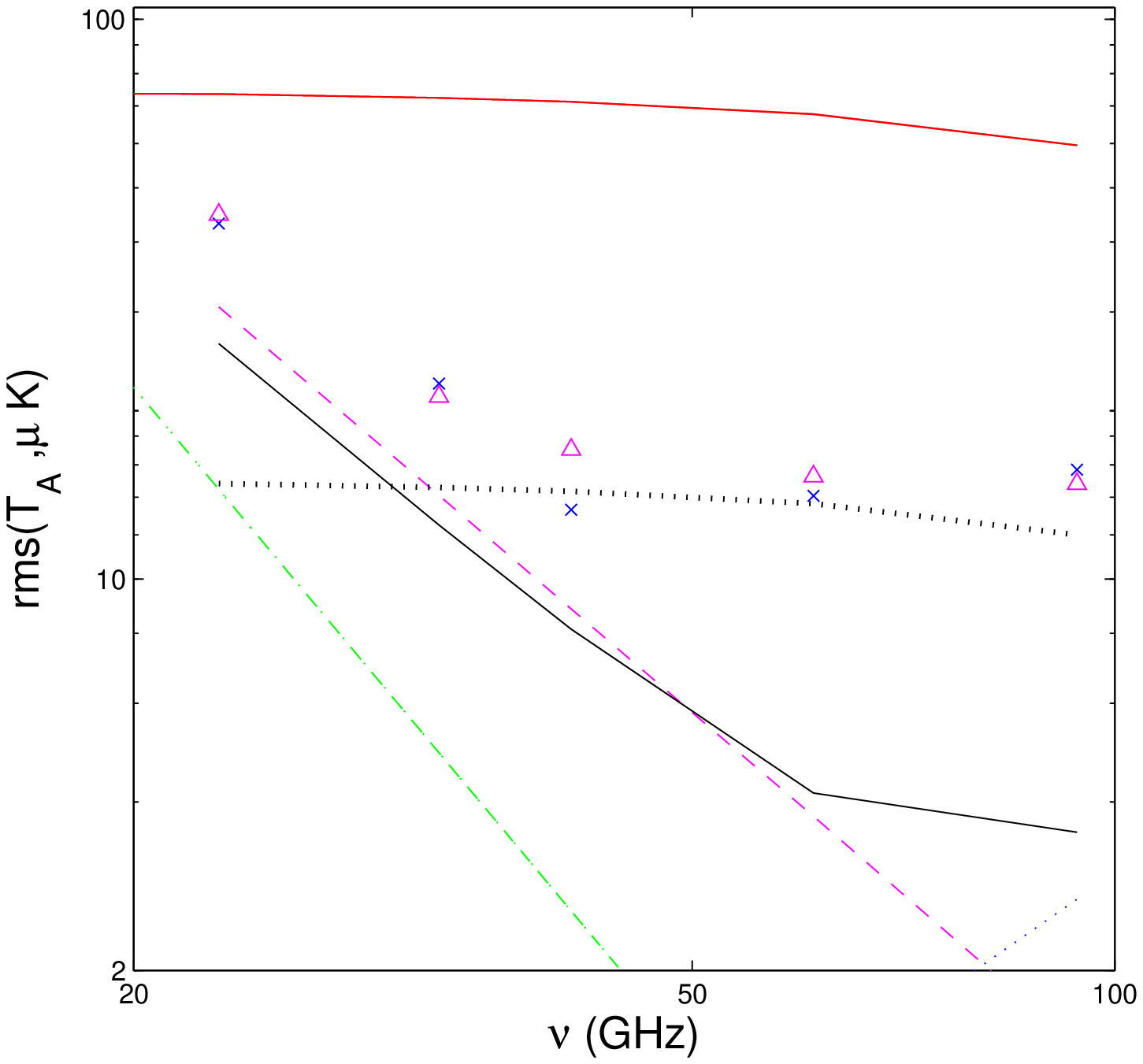}
\caption{Average temperature (left plot) and rms (right plot) in \emph{the Spot} region, as function of frequency. {\bf Foregrounds}: synchrotron
(dot--dashed line), free--free (long--dashed line), dust
(small dots), total Galactic signal (solid dark line).
{\bf CMB} (solid light lines). 
{\bf Residual maps}: WCM (crosses), TCM (triangles). {\bf Noise} (big dots).}
\label{f1}
\end{figure*}
Although the non--Gaussian deviations described in V04 and C05 are only observed
in wavelet space, our starting point is to study foreground behaviours
of \emph{the Spot} region in real space. 
In C05 most important features found in real space were four prominent cold spots with an amplitude higher than 2 times the dispersion of the map.
Their size, amplitude and position can be found in Table 4 of C05. Several tests have
been performed in order to find non--Gaussian features of \emph{the Spot}, considering the total cold area in the region and the four
resolved spots. However, all the tests give results compatible with
Gaussianity. Neither the total area in the region nor the four resolved spots are particularly cold or big, compared to the simulations.

Fig. \ref{fig:SpotImage} shows
a $22\degr\times22\degr$ area of the WCM centered at \emph{the Spot}, as well as the foreground templates scaled to the Q band. 
In the foreground templates we do not observe any significant correlation with the WCM. What can be noticed is that all three foreground
emissions are low in the region of \emph{the Spot}. This means that after the convolution with the SMHW the foreground contribution will be
negative in this region.

In Fig. \ref{f1} we plot the expected mean and rms contribution of
Galactic foregrounds at the five WMAP frequencies in \emph{the Spot} region
(extrapolating the templates as described above), compared to the WCM and TCM CMB
estimates. 
The values in Fig. \ref{f1} are in antenna temperature and the region of \emph{the Spot} is defined by a circle of $10\degr$ radius.
If we focus on the total intensity (left plot), the 
synchrotron emission (dot--dashed lines) is clearly the dominant
Galactic component at K and Ka bands, and it is still higher than the
free--free emission (dashed lines) in the Q band. On the contrary,
when the rms amplitude of the signal is taken into account, the
free--free emission becomes dominant respect to the synchrotron at all
the WMAP frequencies. 
This means that in \emph{the Spot} region, synchrotron
emission is brighter than free--free but much more uniformly
distributed. 
Since in wavelet space uniform components are erased, we expect the free--free radiation to be the dominant foreground in wavelet space.
Anyway, compared to the total rms
signal in WMAP maps, the Galactic foreground contribution seems to be
relevant only at K and Ka bands, whereas at higher frequencies its rms
amplitude is always lower than the WMAP rms noise and at least one
order of magnitude lower than CMB. The contribution of dust emission,
according to the FDS model, is very small, also at 94\,GHz where
it is the most relevant Galactic foreground.

In order to compare with the CMB maps, 
we subtract the WCM and the TCM from the not--cleaned WMAP data at each channel:
the resulting maps should include only foregrounds plus noise. The
two residual maps provide nearly the same average and rms value at all
the frequencies, in a very good agreement with our foregrounds (or
foregrounds+noise) estimate. This is particularly evident at the low
bands K and Ka, where Galactic foregrounds are more relevant. This is
not surprising for the WCM case, because it is based on the same templates
for the dust and free--free emission as our estimation (although the synchrotron template is different). 
On the other hand, the agreement with the TCM map confirms
that the extrapolation of foreground templates is quite reliable. Therefore,
WCM and TCM methods seem to provide a good estimate of CMB in \emph{the Spot}
region in spite of their completely different technique to subtract
foregrounds.

\begin{figure}
\includegraphics[width=84mm]{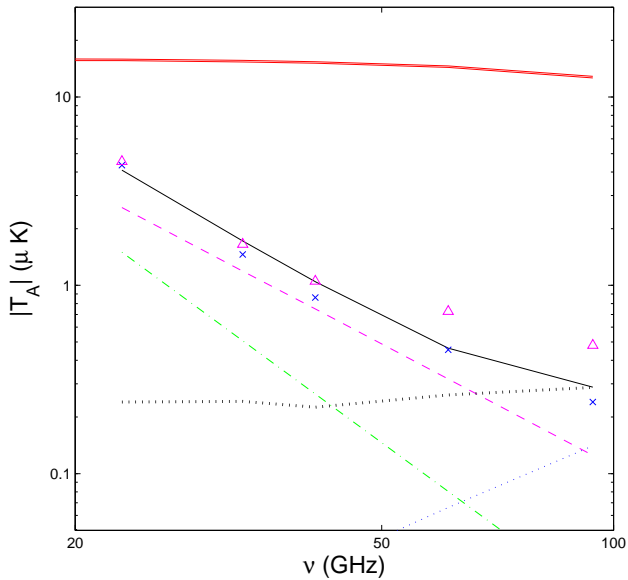}
\caption{As in Fig. \ref{f1} but for the temperature at the center of
\emph{the Spot} in wavelet space.}
\label{f2}
\end{figure}

\subsubsection{The Spot in wavelet space}
\label{ss2}

We know that wavelets are able to amplify hidden non--Gaussian features, lowering  
the noise and removing constants. What can be deduced from the results
of V04 and C05 is that there is a spatial temperature variation in
this region which matches surprisingly well the shape of the SMHW,
increasing its relative signal respect to foregrounds and noise in
wavelet space.

The contribution of foregrounds, already very low in the real space, is
still reduced in wavelet space: it can be appreciated in Fig.
\ref{f2}, where we plot the antenna temperature at the center of \emph{the Spot} at scale $R_8$. 

Note that we plot absolute values because the CMB and foregrounds temperatures are negative at the center of the Spot. 
In wavelet space the foregrounds can show negative temperatures since the convolution with compensated wavelets removes the zero level. 
This means that also the foreground emissions at the center of this region are lower than their average emission in the region 
(see Fig. \ref{fig:SpotImage}).
Comparing the mean temperature in real space with the temperature at the center
of \emph{the Spot} in wavelet space, we find that, while the CMB value is
reduced by a factor $\sim5$, the foreground signal is now a factor
between 7 and 12 lower. This is a confirmation of goodness of an
analysis in wavelet space. Also the noise is strongly reduced and is
always lower than the expected foreground signal, which is clearly
dominated by free--free emission. Finally, as in previous figures, we
show the residual temperature after subtraction of different CMB maps
from WMAP data. They are in a good agreement with our foregrounds estimation. 
\begin{figure*}
\begin{center}
\includegraphics[width=14cm]{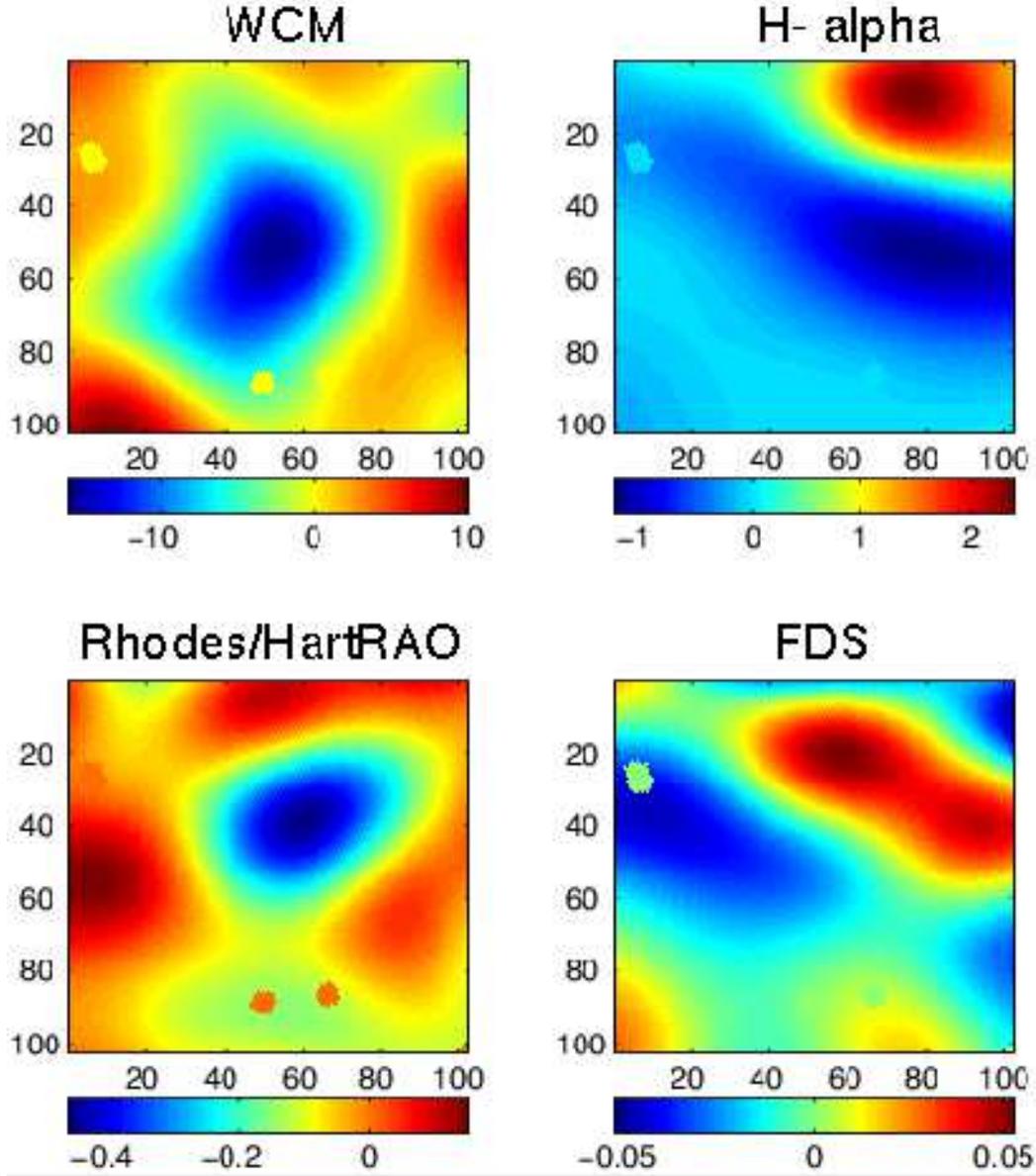}
\end{center}
\caption{In this figure we show a $22\degr\times22\degr$ square centered at the minimum of \emph{the Spot}, for different maps at scale $R_{8}=4.17\degr$.
The same maps as in Fig.\ref{fig:SpotImage} are shown, again scaled to channel Q, in the same units. 
Only the weak synchrotron emission shows some correlation with \emph{the Spot}, but its amplitude is two orders of magnitude lower than \emph{the Spot}.
Therefore we do not expect it to be responsible for the observed deviation from Gaussianity. Note that in all figures we see three small spots which correspond
to masked point sources.}
\label{fig:template_images}
\end{figure*}

The region of \emph{the Spot} in the WCM is
shown in Fig. \ref{fig:template_images} as it appears in wavelet space at
scale $R_8$. Moreover, we report also
the images of foreground templates, scaled to the Q band.
The distribution of foregrounds temperature in wavelet space does not
show any correlation with \emph{the Spot} in CMB maps, although the center of
this region corresponds always to a negative value. The only foreground
whose spatial distribution resembles in some way the CMB spot is
the synchrotron emission. 
However it is very unlikely that a residue of a signal which is two orders of magnitude lower than the total signal could be responsible for the
observed deviation from Gaussianity.

\subsubsection{Can Galactic residues explain the non--Gaussian
Spot in wavelet space?}
\label{ss3}

Even if it seems unlikely from the previous sections, we investigate now the possibility that a foreground
residue in CMB maps could explain the non--Gaussian signal found in
wavelet space by V04 and C05. These papers performed several tests in
order to investigate if foregrounds can affect the results, but no
evidence of it has been found out. On the contrary, different
arguments seem to discard foregrounds as sources of the non--Gaussian
signal in WMAP CMB data: 1) the most significant one is probably the
lack of frequency dependence both in the kurtosis and in properties of
\emph{the Spot}; 2) \emph{the Spot} is located in a region with very low foreground
emission, and their relative contribution is reduced when we go from
real space (where no deviations from Gaussianity are found) to wavelet
space; 3) similar results are found using totally independent techniques
to subtract foregrounds (e.g., WCM and TCM maps; see next paragraphs).
Let us see now more details for the different involved estimators.

\begin{figure*}
\includegraphics[width=84mm]{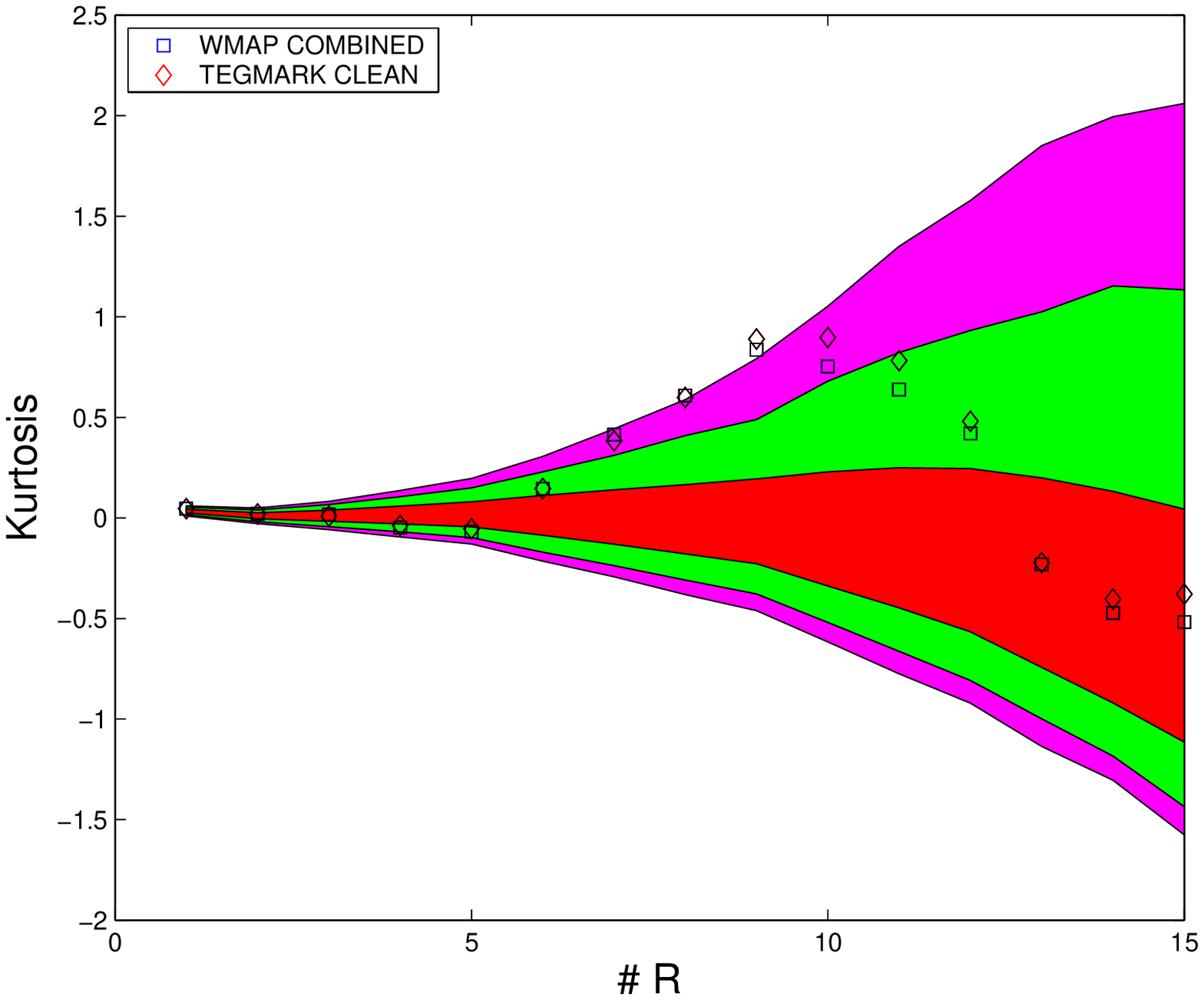}
\includegraphics[width=84mm]{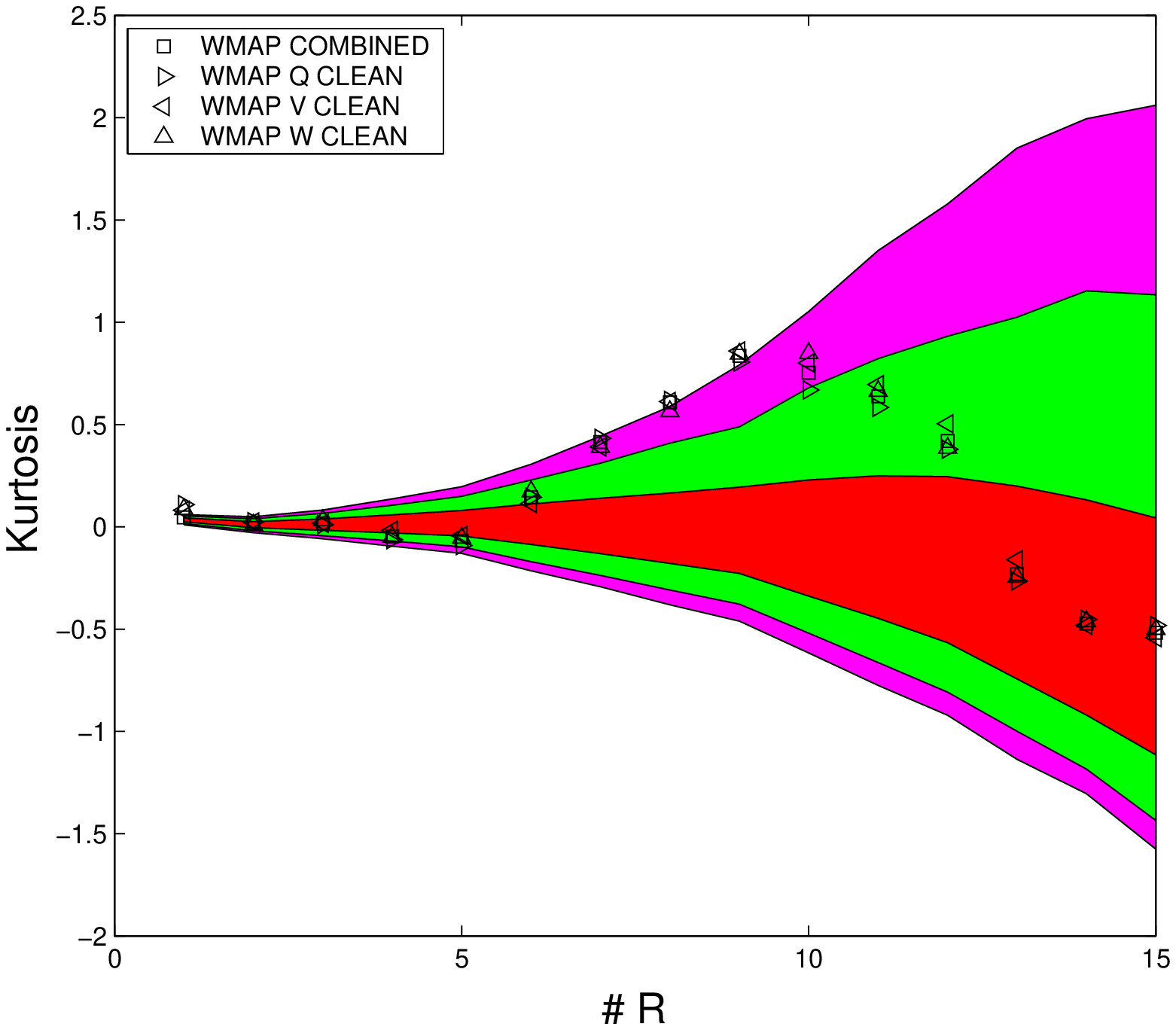}
\caption{Comparison between kurtosis found in WCM and TCM clean maps (left plot). 
  In the right plot we compare the kurtosis values of the Q, V and W bands with the WCM. 
  The acceptance intervals
  for the 32\% (inner), 5\% (middle) and 1\% (outer) significance levels,
  given by the 10000 simulations are also plotted.
  Note that the figure on the right is similar to Fig. 7 in V04, but the Q map values have been corrected since they were slightly different
  due to an error in the construction of the Q map.}
\label{f3}
\end{figure*}

{\bf Kurtosis}: this statistic has been deeply investigated in V04. Here we want only to stress some points. Contrary to what is
expected if foregrounds contamination is relevant, the distribution of
kurtosis as function of wavelet scales has the same pattern and
amplitude in the WCM as well as in each single
band Q, V and W (see Fig.\ref{f3}, right plot). Moreover, a similar shape for the
kurtosis distribution has been obtained from the TCM map (see Fig.
\ref{f3}, left plot). The only relevant difference is noticed at the peak of the
distribution, where the amplitude of the kurtosis is higher, meaning an even  
more significant deviation from Gaussianity.

{\bf Temperature at the center of the Spot}: in wavelet space at scale
$R_8$, \emph{the Spot} presents an extremely negative temperature in its
minimum.
Considering the WCM, the minimum has a
temperature of $\sim-16.1\mu$K, nearly 4.6 times the dispersion of
that map. Considering $10000$ Gaussian CMB simulations, the probability of finding a minimum of such amplitude or less at scale $R_8$, is $0.75\%$.
No particular frequency dependence has been observed. In Table
\ref{table:amplitude} the temperature at the center of \emph{the Spot} is
also reported for the combined not--cleaned WMAP map, (hereafter WNCM), and for maps
cleaned by different techniques.
\begin{table}
  \begin{center}
    \caption{Temperature at the center of \emph{the Spot} for different maps at
      scale $R_8$. The values are expressed in thermodynamic
      temperature and in terms of the dispersion of the corresponding
      map. The last column gives the probability of having a lower or equal minimum temperature under the Gaussian hypothesis. 
      The ``1$\%$ limit'' is given by the simulation whose minimum temperature has 1$\%$ upper tail probability compared 
      to the 10000 Gaussian simulations.} 	  
    \begin{tabular}{|c|c|c|c|c|}
      \hline
      Map & $T(\mu$K) & $\sigma (\mu$K) & $n_{\sigma}=T/\sigma$ & upper tail \\
          &&&&                                                    probability\\
      \hline
      1$\%$ limit & & & -4.52 & 1.00$\%$  \\
      \hline
      WNCM & -16.39 & 3.55  & -4.62 & 0.61$\%$ \\
      \hline
      WCM & -16.12 & 3.52 & -4.58 & 0.75$\%$ \\
      TCM & -15.84 & 3.47 & -4.56 & 0.82$\%$ \\
      \hline
      ILC & -16.30 & 3.54 & -4.61 & 0.63$\%$ \\
      WFM & -13.87 & 3.01 & -4.61 & 0.61$\%$ \\
      MEM map & -16.48 & 3.53 & -4.67 & 0.44$\%$ \\
      Local subtraction & -15.63 & 3.52 & -4.44 & 1.32$\%$ \\
      \hline
    \end{tabular}
    \label{table:amplitude}
  \end{center}
\end{table}
The WNCM is obtained by combining the not cleaned data in the same way as the WCM.
We estimate the significance of the
deviation from Gaussianity and we find that it is always below the $1\%$ upper tail probability,
independently of the technique used
to subtract Galactic foregrounds. 

Nevertheless, the difference in
temperatures between the observed temperature and the one giving an upper tail probability of $1\%$ are
$\approx0.2\mu$K for TCM and WCM maps, lower or of the same order than
noise and foregrounds contribution (see Fig. \ref{f2}). We can
conclude that this test gives a less robust non--Gaussian detection 
compared to kurtosis and area tests. For example, subtracting our
Galactic signal estimate in \emph{the Spot} region, from the WNCM, we
see that the upper tail probability is slightly above $1\%$. 
As in this case we do not know the $\sigma$ of the cleaned map, since our estimation is local, we use the $\sigma$ of the WCM map
(see the last row of Table \ref{table:amplitude}).

{\bf Area of the Spot}: a stronger deviation from Gaussianity is
observed from the dimension of \emph{the Spot} area. Considering only pixels
with lower temperatures than 3 times the dispersion of the map, \emph{the Spot}
covers a region of the sky of diameter $\approx8\degr$ at wavelet scale $R_8$. Based on the WCM map, C05
showed that the probability of having a spot as big as this one in
Gaussian simulations is always smaller than $0.65\%$ for thresholds
between 3 and 4.5$\sigma$. Moreover, the area of \emph{the Spot} has a similar
dimension at all the WMAP frequencies. 
\begin{figure}
\includegraphics[width=84mm]{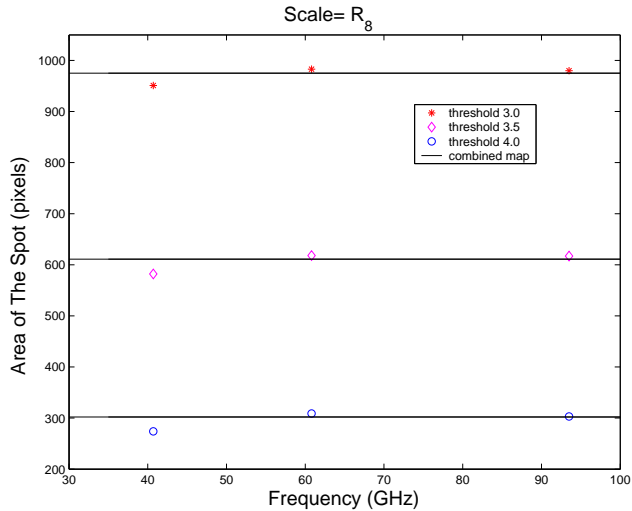}
\caption{Frequency dependence of the area of \emph{the Spot}, at thresholds 3.0 (asterisks), 3.5 (diamonds) and 4.0 (circles), at scale $R_8$. 
  The horizontal lines represent the combined map values. The area is almost frequency independent. Note that this figure is 
  similar to Fig. 11 in C05, but the Q map values have been corrected since they were slightly lower due to an error in the construction of the Q map.}
\label{fig:Area_freq}
\end{figure}
This can be observed in Fig. \ref{fig:Area_freq}.
We report in Table \ref{table:area} the dimension
in pixels of \emph{the Spot} area in the different clean maps with the
corresponding upper tail probability: for all the considered maps, 
this probability is around 0.2--0.4$\%$. Contrary to the
minimum temperature estimator, if we subtract our local foreground estimate 
from the not cleaned maps, the upper tail probability of the detection still remains
widely below 1$\%$. Even if two times the estimate is subtracted,
the upper tail probability is still around 1$\%$.
However the strong frequency dependence of this map in Q--V--W bands is incompatible with the observations
(see Table \ref{table:area2}). If we try to explain \emph{the Spot} as a combination of CMB plus bad subtracted foregrounds, the frequency dependence 
must be flat once the contaminating foregrounds are subtracted.

Finally, we discuss the possibility of having a residual foreground contamination
in the clean WMAP maps that is constant at the Q, V and W bands.
Because of their opposite spectral behaviour, we can find a linear
combination of free--free and dust emission with a small variation
between 40 and 94\,GHz. We consider the possibility that the foreground amplitude
estimated from the templates is significantly underestimated. 
Assuming free--free to be underestimated by a 20$\%$ and dust by a 120$\%$, we obtain a nearly constant residue. In this case, the total residue
would be 0.19, 0.14 and 0.20$\mu$K at 41, 61 and 94\,GHz respectively.
These values are not incompatible with possible errors in templates
and in their extrapolation to microwave frequencies (Dickinson, Davies \& Davis 2003;
Finkbeiner et al. 1999). Now, our local foregrounds estimate plus the
constant residue is subtracted from the unclean spot region.
Results are reported in Table \ref{table:area2}. Even subtracting two times the flat residue, we do not obtain any relevant reduction in the 
significance of the non--Gaussian detection(see last two rows in Table \ref{table:area2}).

\begin{table}
  \begin{center}
    \caption{Area and significance of \emph{the Spot} for different maps at
      scale $R_8$. The areas are the number of pixels colder than three times the standard deviation of the corresponding map.
      The first row of the table stands for the simulation whose biggest spot presents a 1$\%$ upper tail probability compared to the biggest spot of 
      each of the 10000 simulations.}
    \begin{tabular}{|c|c|c|c|}
      \hline
      Map & $\sigma (\mu$K) & $N_{pixels}(T>3\,\sigma)$ & upper tail \\
          &&&                                             probability\\
      \hline
      1$\%$ limit & & 831 & 1.00$\%$ \\
      \hline
      WNCM & 3.55  & 990 & 0.32$\%$ \\
      \hline
      WCM & 3.52 & 975 & 0.34$\%$ \\
      TCM & 3.48 & 970 & 0.36$\%$ \\
      \hline
      ILC & 3.54 & 1023 & 0.22$\%$ \\
      WFM & 3.01 & 984 & 0.33$\%$ \\
      MEM & 3.53 & 1011 & 0.26$\%$ \\
      Local subtr. & 3.519 & 924 & 0.45$\%$ \\
      \hline
    \end{tabular}
    \label{table:area}
  \end{center}
\end{table}

\begin{table}
  \begin{center}
    \caption{Area of \emph{the Spot} (in pixels) after local subtraction of foregrounds. 
      The only time where the significance of the
      area can be reduced below the 1$\%$ limit (831 pixels) occurs subtracting two times the local foreground estimation at the Q band. 
      However the remaining spot shows a clear frequency dependence which can not correspond to a clean CMB spot. The other three cases
      show an almost flat cleaned spot, but its size is higher than the 1$\%$ limit given by the Gaussian simulations.}
    \begin{tabular}{|l|c|c|c|}
      \hline
      Map/Band & Q & V & W \\
      \hline
      Local subtraction & 909 & 947 & 949 \\
      $2\times$Local subtraction & 763 & 886 & 909 \\
      Local $+$ flat residue subtr. & 878 & 926 & 923 \\
      Local $+2\times$flat residue subtr. & 860 & 915 & 903 \\
      \hline
    \end{tabular}
    \label{table:area2}
  \end{center}
\end{table}

\section{Conclusions}

In this paper we address the issue of the origin of non--Gaussian behaviours observed in the WMAP data. In particular, a
non--zero kurtosis in the distribution of wavelet coefficients was 
detected by V04 at angular scales ranging from $3\degr$ and $5\degr$. This non--Gaussian signal is mainly generated by the
presence of a very cold spot in the southern
hemisphere, at Galactic coordinates $b=-57\degr$ and $l=209\degr$
(V04, C05). Its dimension ($\approx8\degr$) and temperature in wavelet space makes
this spot quite exceptional compared to Gaussian CMB simulations:
less than $1\%$ of the simulations have spots with similar
characteristics.

As a first step, we have verified the robustness of the deviation from Gaussianity in the kurtosis. We have performed a test taking into account
the estimators used in that detection (skewness and kurtosis) and the
number of consecutive wavelet scales presenting a significant deviation from Gaussianity, namely 4 considering
only the southern hemisphere. The significance for the non--Gaussian
detection remains still high: we obtained that only $0.69\%$
of the simulations have equal or higher deviation of skewness or
kurtosis at any four consecutive scales and in any of the Galactic
hemispheres. 

Afterwards we studied the morphology of \emph{the Spot} using Elliptical Mexican
Hat Wavelets on the sphere. We observed that the maximum amplification of \emph{the Spot}
temperature is obtained for almost isotropic Mexican Hat Wavelets,
meaning that the shape of the underlying signal is essentially circular. 
This result does not discard for instance topological defects like textures or a gravitational potential with a circular shape as possible explanations. 

Finally, we focus on the possible foreground contamination of \emph{the Spot} region in the clean WMAP maps, considering the SZ effect or 
bad--subtracted Galactic foregrounds.
The SZ effect is clearly discarded by the flat frequency dependence of \emph{the Spot} temperature. 
The Galactic foregrounds case requires a more detailed analysis, since a hypothetical foreground mixing could provide a flat foreground contribution. 

In wavelet space, the contribution of Galactic foregrounds in the region of \emph{the Spot} is extremely
low, at least one order of magnitude less than CMB at the Q, V and W
bands. The dominating foreground is the free--free emission at all WMAP
frequencies except at the W band where free--free is at the same level as dust emission and below the noise level.  

If the non--Gaussian analysis is affected by unsubtracted Galactic foregrounds, we would expect that the
non--Gaussian detection is more significant at frequencies where
the foreground emission is more relevant. But neither the kurtosis nor the area and amplitude of \emph{the Spot} are more
significant at the Q band, i.e. the band where the Galactic contribution is higher respect to the V and W bands.
In addition we obtain very similar results from CMB maps produced by completely independent foregrounds subtraction techniques (e.g. WCM and TCM).

Because of the large uncertainties in Galactic emission at microwave
frequencies, we have even considered the possibility that our
templates provide an important underestimate of foregrounds. 
Nevertheless, the possibility of having a strong and frequency independent foreground residue, which could explain the non-Gaussian 
nature of \emph{the Spot}, is very unlikely.

According to our knowledge on Galactic emissions, we can conclude that there is no evidence for a relevant contribution of
unsubtracted foregrounds in the region of the sky which is responsible for the non--Gaussian detection in wavelet space.

\section*{acknowledgments}
Authors kindly thank L.Cay\'on, E.Komatsu and H.K.K. Eriksen for very useful comments and J.L. Jonas for providing us the Rhodes/HartRAO 2326--MHz 
radio survey data. 
MC thanks Spanish Ministerio de Educacion Cultura y Deporte (MECD) for a predoctoral FPU fellowship.
We acknowledge partial financial support from the Spanish MCYT project ESP2004-07067-C03-01 and 
the use of the Legacy Archive for Microwave Background Data 
Analysis (LAMBDA). Support for LAMBDA is provided by the NASA Office of Space 
Science.
This work has used the software package HEALPix (Hierarchical, Equal
Area and iso-latitude pixelization of the sphere,
http://www.eso.org/science/healpix), developed by K.M. G{\'o}rski,
E. F. Hivon, B. D. Wandelt, J. Banday, F. K. Hansen and
M. Barthelmann and the visualisation program Univiewer, developed by S.M. Mingaliev, M. Ashdown and V. Stolyarov. We acknowledge the help 
of V. Stolyarov related to the instalation and usage of this program.

\end{document}
\end